\documentclass[aps,prl,twocolumn,superscriptaddress,showpacs]{revtex4}
\usepackage{amsmath}
\usepackage{amssymb}
\usepackage{tikz}
\usepackage{graphicx}
\usepackage{dcolumn}
\usepackage{bm}
\usepackage{epsfig}
\usepackage{psfrag}

\newcommand{\reffig}[1]{\mbox{Fig.~\ref{#1}}}

\newcommand{\T}{${\mathcal T}\,$}
\newcommand{\Ti}{${\mathcal T}$}
\begin{document}
\title{\bf  Power spectrum analysis and missing level statistics of microwave graphs with violated time reversal invariance}

\author{Ma{\l}gorzata Bia{\l}ous}
\affiliation{Institute of Physics, Polish Academy of Sciences, Al. Lotnik\'ow 32/46, 02-668 Warszawa, Poland}
\author{Vitalii Yunko}
\affiliation{Institute of Physics, Polish Academy of Sciences, Al. Lotnik\'ow 32/46, 02-668 Warszawa, Poland}
\author{Szymon Bauch}
\affiliation{Institute of Physics, Polish Academy of Sciences, Al. Lotnik\'ow 32/46, 02-668 Warszawa, Poland}
\author{Micha{\l} {\L}awniczak}
\affiliation{Institute of Physics, Polish Academy of Sciences, Al. Lotnik\'ow 32/46, 02-668 Warszawa, Poland}
\author{Barbara Dietz}
\email{dietz@ifpan.edu.pl}
\affiliation{Institute of Physics, Polish Academy of Sciences, Al. Lotnik\'ow 32/46, 02-668 Warszawa, Poland}
\author{Leszek Sirko}
\email{sirko@ifpan.edu.pl}
\affiliation{Institute of Physics, Polish Academy of Sciences, Al. Lotnik\'ow 32/46, 02-668 Warszawa, Poland}

\date{\today}

\bigskip

\begin{abstract}

We present experimental studies of the power spectrum and other fluctuation properties in the spectra of microwave networks simulating chaotic quantum graphs with violated time reversal invariance. On the basis of our data sets we demonstrate that the power spectrum in combination with other long-range and also short-range spectral fluctuations provides a powerful tool for the identification of the symmetries and the determination of the fraction of missing levels. Such a procedure is indispensable for the evaluation of the fluctuation properties in the spectra of real physical systems like, e.g., nuclei or molecules, where one has to deal with the problem of missing levels.   
\end{abstract}

\bigskip
\maketitle

{\it Introduction.}--- 
In the last decades the concept of quantum chaos, that is, the understanding of the features of the classical dynamics in terms of the spectral properties of the corresponding quantum system, like nuclei, atoms, molecules, quantum wires and dots or other complex systems~\cite{Gomez2011,Weidenmueller2009,Haake2001}, has been elaborated extensively.  It has been established by now that the spectral properties of generic quantum systems with classically regular dynamics agree with those of Poissonian random numbers~\cite{Berry1977} while they coincide with those of the eigenvalues of random matrices~\cite{Mehta1990} from the Gaussian orthogonal ensemble (GOE) and the Gaussian unitary ensemble (GUE) for classically chaotic systems with and without time-reversal (\Ti) invariance~\cite{Footnote}, respectively, in accordance with the Bohigas-Giannoni-Schmit (BGS) conjecture~\cite{Bohigas1984}. 

A multitude of studies with focus on problems from the field of quantum chaos have been performed by now theoretically and numerically. However, there are nongeneric features in the spectra of real physical systems that are not yet fully understood. Such problems are best tackled experimentally with the help of model systems like microwave billiards~\cite{Stoeckmann2000,Dietz2015} and microwave graphs~\cite{Hul2004,Lawniczak2010}. In the experiments with microwave billiards the analogy between the scalar Helmholtz equation and the Schr\"odinger equation of the corresponding quantum billiard is exploited. Microwave graphs~\cite{Hul2004,Lawniczak2010} simulate the spectral properties of quantum graphs~\cite{Kottos1997,Kottos1999,Pakonski2001}, networks of one-dimensional wires joined at vertices. They provide an extremely rich system for the experimental and the theoretical study of quantum systems, that exhibit a chaotic dynamics in the classical limit. 

The idea of quantum graphs was introduced by Linus Pauling to model organic molecules \cite{Pauling1936} and they are also used to simulate, e.g., quantum wires~\cite{Sanchez1988}, optical waveguides~\cite{Mittra1971} and mesoscopic quantum systems~\cite{Kowal1990,Imry1996}. The validity of the BGS conjecture was proven rigourously for graphs with incommensurable bond lengths in Refs.~\cite{Gnutzmann2004,Pluhar2014}. Accordingly, the fluctuation properties in the spectra of classically chaotic quantum graphs with and without \T invariance are expected to coincide with those of random matrices from the GOE and the GUE, respectively. This was confirmed experimentally~\cite{Hul2004,Lawniczak2010} for the nearest-neighbor spacing distribution using microwave networks~\cite{Lawniczak2008,Lawniczak2011,Hul2012,Lawniczak2014,Allgaier2014}. 

The statistical analysis of the spectral properties of a quantum system and the comparison with the conventional GOE or GUE results requires complete sequences of eigenvalues belonging to the same symmetry class~\cite{Bohigas1983,Bohigas1984}. Accordingly, the experimental determination of the chaoticity of a system on the basis of the spectral fluctuation properties might be far from simple, since several effects, like, e.g., nongeneric contributions as in the case of the stadium billiard~\cite{Sieber1993}, the existence of tiny islands of regular dynamics in the chaotic sea~\cite{Dietz2014}, mixed symmetries or incomplete spectra may result in deviations from the random-matrix theory (RMT) predictions. 

We are not aware of experimental studies including the analysis of long-range spectral fluctuations in incomplete spectra of chaotic systems with violated \T invariance, which, as outlined below, is essential to be able to obtain conclusive results on the spectral properties. Our objective is to fill this gap. \T violation was tested experimentally, e.g., in nuclear spectra and in compound-nucleus reactions~\cite{French1985,Mitchell2010} and in electron transport through quantum dots, where \T violation is induced by a magnetic field~\cite{Pluhar1995}. Furthermore, \T violation in scattering systems was studied thoroughly in experiments with microwave billiards~\cite{Dietz2009,Dietz2010}. The effects of \T violation on the spectral properties of the eigenvalues of closed quantum systems have also been investigated in such systems~\cite{So1995,Stoffregen1995,Wu1998}. However, it is difficult if not impossible to obtain complete \T violation in microwave billiards, whereas its achievement is straightforward in microwave networks~\cite{Lawniczak2008,Lawniczak2011,Hul2012,Lawniczak2014,Allgaier2014}. 

In this Letter we will develop a procedure to obtain information on the chaoticity and \T symmetry of a classical system from the spectral properties of the corresponding quantum system in the presence of missing levels. Incomplete spectra are actually a problem one has to cope with in real physical systems like, e.g.,  nuclei and molecules~\cite{Liou1972,Zimmermann1988,Frisch2014,Mur2015}, so such a procedure is a requisite for their analysis~\cite{Agvaanluvsan2003,Bohigas2004}. It is applied to the spectra of irregular, fully connected microwave networks simulating quantum graphs with violated \T invariance. The impact of missing levels on the spectral fluctuation properties is particularly large for long-range spectral fluctuations. It was demonstrated numerically in Ref.~\cite{Molina2007} that the power spectrum~\cite{Relano2002,Faleiro2004} is a powerful statistical measure to discriminate between deviations caused by missing levels and by the mixing of symmetries. Additional evidence for these effects may be obtained on the basis of commonly used statistical measures for short- and long-range spectral fluctuations~\cite{Bohigas2004}. Accordingly, in order to unambiguously identify the symmetry of the system and the fraction of missing levels, we considered all these statistical measures.    

\begin{figure}[h!]
\includegraphics[width=0.8\linewidth]{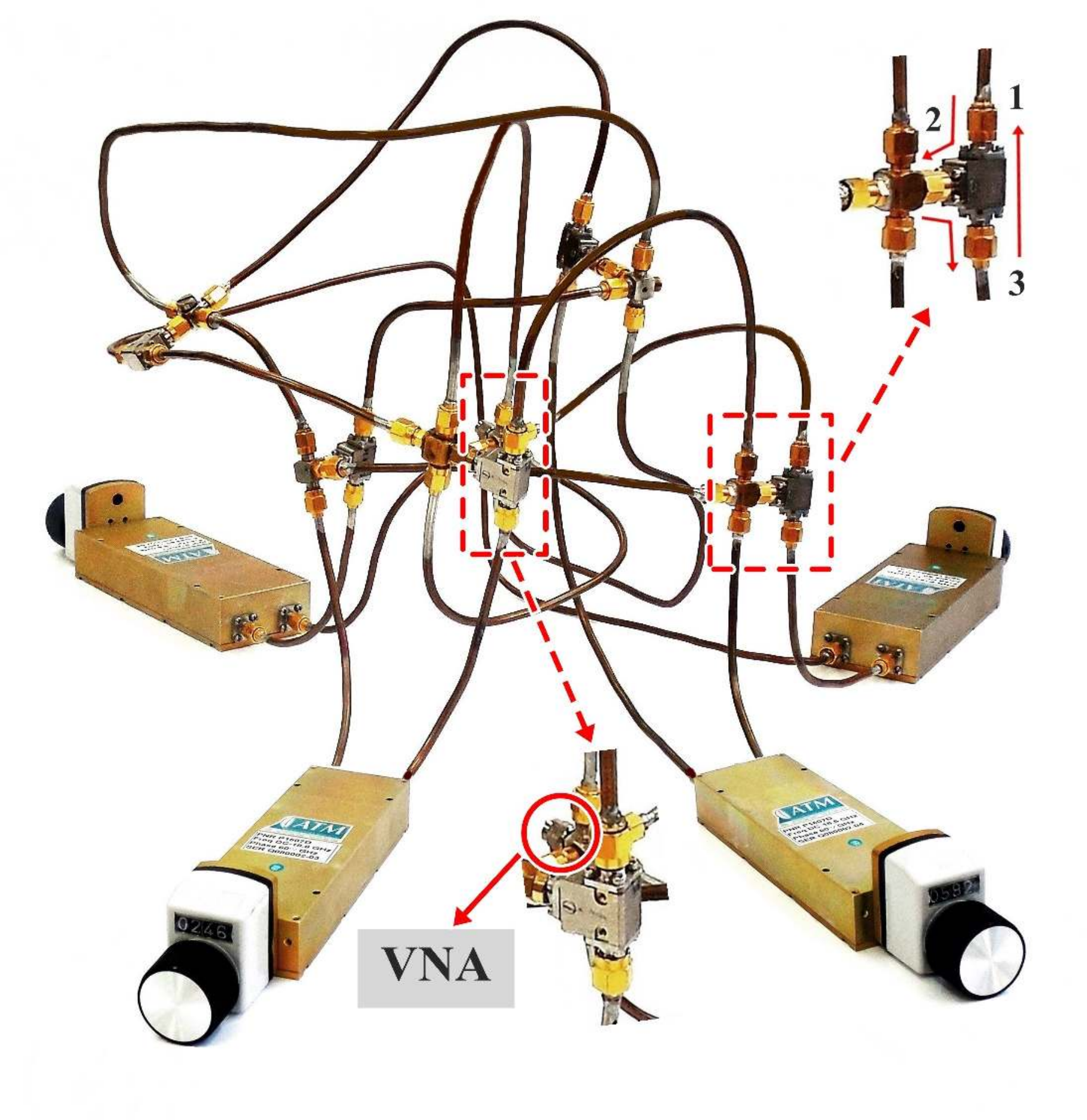}
\caption{
Photograph of one realization of a microwave network. An ensemble of 30 different networks was created by changing the lengths of four bonds using the phase shifters visible at the bottom of the graph. Time-reversal invariance was induced by five microwave circulators. One is shown enlarged in the right-upper inset to illustrate their functionality. For the measurement of the scattering matrix, the vector network analyzer (VNA) was coupled to the network via a HP 85133-616 flexible microwave cable; see lower inset.}
\label{Fig1}
\end{figure}

{\it Experimental setup.}--- 
We simulate quantum graphs experimentally by using a network of coaxial microwave cables, that are coupled by junctions at the vertices. A photograph of one example is shown in Fig.~\ref{Fig1}. The microwave networks comprised 6 junctions, that were all connected with each other by coaxial cables, in order to simulate a fully connected quantum graph. The coaxial cables (SMA-RG402) consist of an inner conductor of radius $r_1=0.05$~cm, which was surrounded by a concentric conductor of inner radius $r_2=0.15$~cm. The space between them was filled with Teflon. Measurements yielded a dielectric constant $\varepsilon\simeq 2.06$. Below the cut-off frequency of the TE$_{11}$ mode $\nu_{c}\simeq\frac{c}{\pi (r_1+r_2)\sqrt{\varepsilon}} = 33.26$~GHz~\cite{Jones,Savytskyy2001} only the fundamental TEM mode can propagate inside a coaxial cable. Note, that not the geometric lengths $L_i$ of the coaxial cables, but the optical lengths $L^{opt}_i=L_i\sqrt{\varepsilon}$ yield the lengths of the bonds in the corresponding quantum graph. The analogy between a quantum graph and a microwave network with the same topology relies on the formal equivalence of the wave equations governing the wave function $\psi_{ij} (x)$ of a particle moving in the bond connecting vertices $i$ and $j$ of a quantum graph and the potential difference $U_{ij} (x)$  between the inner and the outer conductors in the corresponding coaxial cable. In the first case, the equation is given by the one-dimensional Schr\"odinger equation with Neumann boundary conditions at the vertices connecting the different bonds. In the second case, it coincides with the Telegraph equation, again with Neumann boundary conditions at the junctions connecting the coaxial cables.

\begin{figure}[h!]
\includegraphics[width=0.9\linewidth]{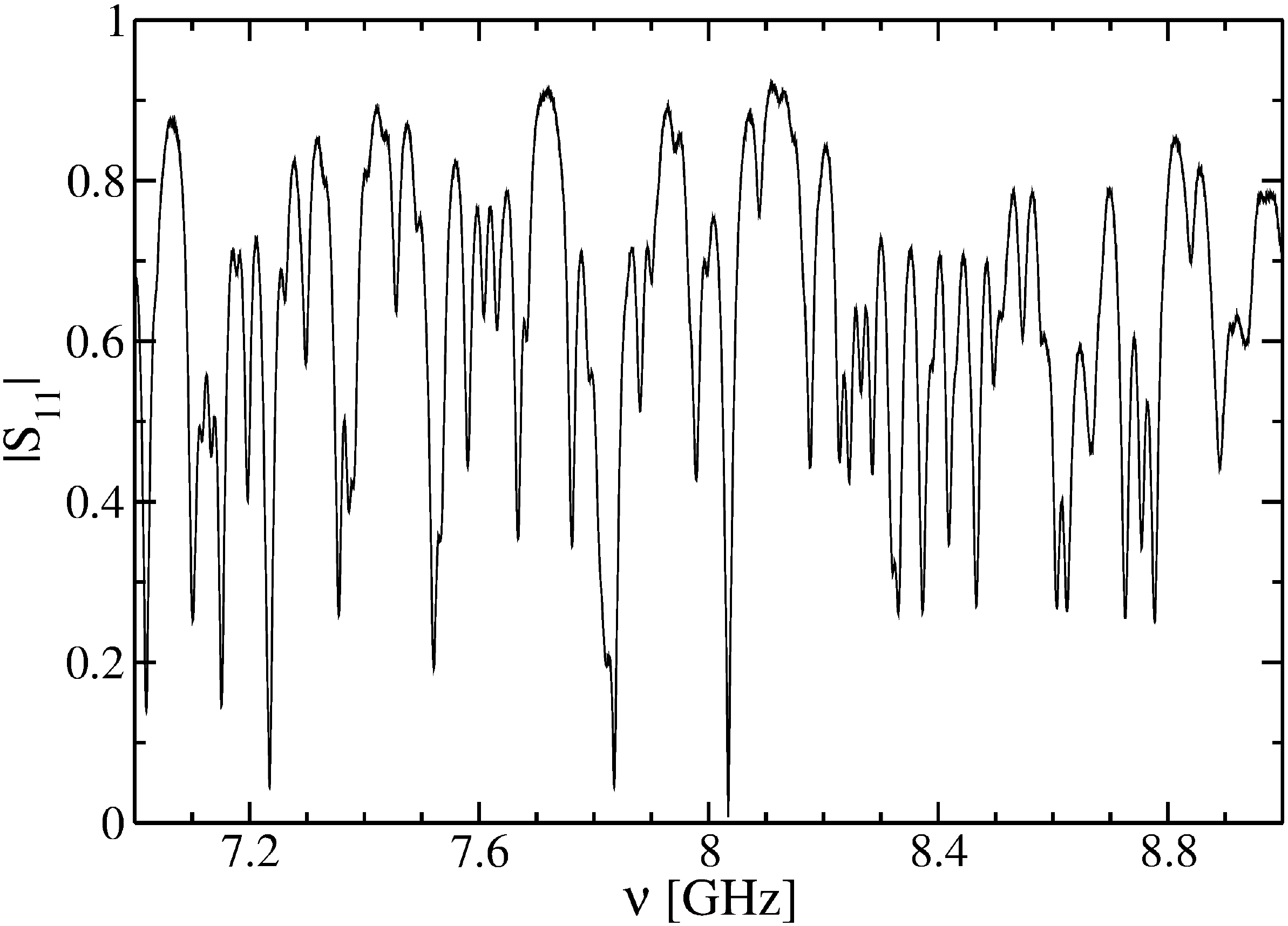}
\caption{
A reflection spectrum starting at 7~GHz. The circulators operate only above that frequency. The resonances obviously overlap. This makes the determination of the resonance frequencies very difficult and thus calls for a reliable theoretical description which accounts for missing levels.
}\label{Fig2}
\end{figure}

The \T violation was induced with five Anritsu PE8403 microwave circulators with low insertion loss which operate in the frequency range from $7-14$~GHz. These are non-reciprocal three-port passive devices. A wave entering the circulator through port 1, 2 or 3 exits at port 2, 3, or 1, respectively, as illustrated schematically in the right-upper inset of Fig.~\ref{Fig1}. The scattering matrix element $S_{11}$ was measured using an Agilent E8364B microwave vector network analyzer (VNA), connected to a six-arm vertex of the network via a HP 85133-616 flexible microwave cable; see lower inset in Fig.~\ref{Fig1}. Figure~\ref{Fig2} shows a part of one measured reflection spectrum. Due to the unavoidable absorption in the walls of the cables used as bonds it exhibits weakly overlapping resonances, of which the positions yield the eigenvalues of the corresponding quantum graph. Accordingly, their determination was a non-trivial task. We compared the measured reflection spectra of an ensemble of 30 different realizations of graphs with the same total optical length $\mathcal{L}\simeq 7.2$~m. It was generated by varying the lengths of four bonds of lengths $L_i\simeq 50-65$~cm with phase shifters (see Fig.~\ref{Fig1}) in steps of $\pm\, 0.112$~cm, thus yielding slightly differing positions of the resonances. An estimate using Weyl's law for quantum graphs~\cite{Kottos1999} indicated that approximately 4~$\%$ of the eigenvalues were missing. 

{\it Fluctuations in the experimental spectra.}---
For the analysis of the spectral properties of the microwave networks, first, the resonance frequencies need to be rescaled in order to eliminate system specific properties like the total length $\mathcal{L}$ of the graph. This is done with the help of Weyl's law, which states, that the resonance density $\rho(\nu)=\mathcal{L}/(2\pi)$ is uniform. Accordingly, the rescaled eigenvalues are determined from the resonance frequencies as $\epsilon_i=\nu_i\mathcal{L}/(2\pi)$, with the frequencies sorted such that $\nu_i\leq\nu_{i+1}$. 

\begin{figure}[h!]
\includegraphics[width=0.9\linewidth]{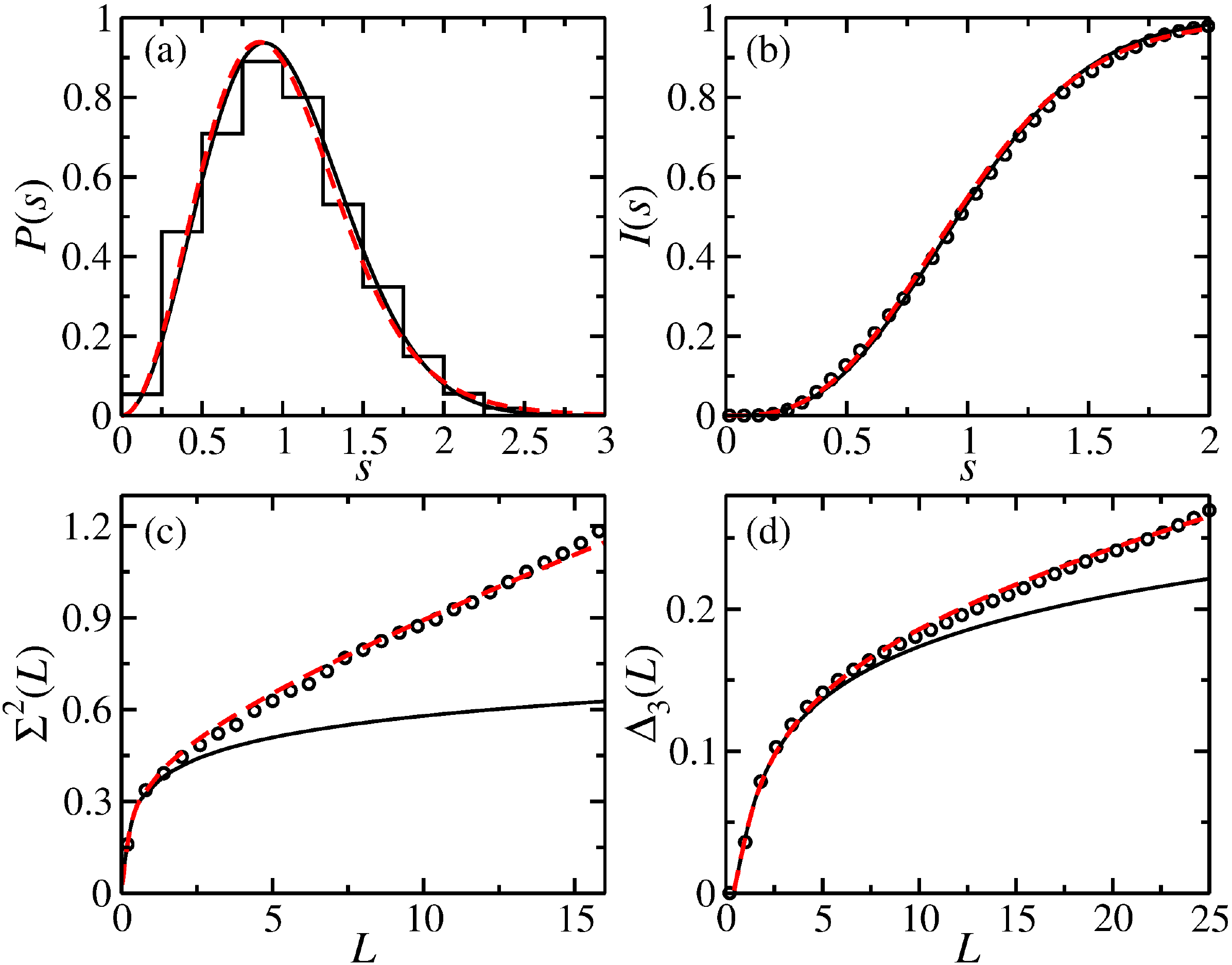}
\caption{(Color online) Spectral properties of the rescaled resonance frequencies. Panels (a)-(d) show the nearest-neighbor spacing distribution $P(s)$ (histogram), its integral $I(s)$ (circles), the number variance $\Sigma^2$ (circles) and the stiffness $\Delta_3$ statistic (circles), respectively. The experimental results are compared to those of the eigenvalues of random matrices from the GUE curves (black full lines) and the corresponding missing-level statistics (red [gray] dashed lines) with $\varphi=0.965$; see main text. 
}\label{Fig3}
\end{figure}

A commonly used measure for short-range spectral fluctuations is the nearest-neighbor spacing distribution, that is, the distribution of the spacings between adjacent eigenvalues, $s_i=\epsilon_{i+1}-\epsilon_i$. For long-range spectral fluctuations these are the variance $\Sigma^2(L)$ of the number of eigenvalues in an interval $L$ and the stiffness of the spectrum $\Delta_3(L)$, given by the least-squares deviation of the integrated resonance density of the eigenvalues from the straight line best fitting it in the interval $L$~\cite{Mehta1990}. The histogram and the circles in Fig.~\ref{Fig3} show the nearest-neighbor spacing distribution $P(s)$ in (a), its integral $I(s)=\int_0^s{\rm d}s^\prime P(s^\prime)$ in (b), the number variance $\Sigma^2$ in (c) and the stiffness $\Delta_3$ in (d). The experimental curves were generated by computing the averages of the statistical measures obtained for each of the 30 microwave networks. Here, for each of them 250 resonance frequencies could be identified. While the short-range spectral fluctuations in (a) and (b) seem to coincide well with those of the eigenvalues of random matrices from the GUE (full black lines), this is not the case for the long-range spectral fluctuations in (c) and (d).

Another statistical measure for long-range spectral fluctuations is the power spectrum of the deviation of the $q$th nearest-neighbor spacing from its mean value $q$, $\delta_q=\epsilon_{q+1}-\epsilon_1-q$. It is given in terms of the Fourier spectrum from 'time' $q$ to $k$, $S(k)=|\tilde{\delta}_k |^2$, with
\begin{equation}
\label{Eq.1}
\tilde{\delta}_k=\frac{1}{\sqrt{N}}\sum_{q=0}^{N-1} \delta_q\exp\left(-\frac{2\pi ikq}{N}\right)
\end{equation}
when considering a sequence of $N$ levels. The power spectrum has not established itself widely, even though, as we will demonstrate in this Letter, it provides a particularly useful statistical measure, especially in the presence of missing levels. It was shown in Refs.~\cite{Relano2002,Faleiro2004}, that for $k/N\ll 1$ the power spectrum which, in fact, only depends on the ratio $\tilde k=k/N$ exhibits a power law dependence $\langle S(\tilde k)\rangle\propto (\tilde k)^{-\alpha}$. Here, for regular systems $\alpha =2$ and for chaotic ones $\alpha =1$ independently of whether \T invariance is preserved or not. The power spectrum and this power law behavior was studied numerically in Ref.~\cite{Robnik2005,Salasnich2005,Santhanam2005,Relano2008}, experimentally in a microwave billiard with classically chaotic dynamics in Ref.~\cite{Faleiro2006} and for a singular rectangular microwave billiard in Ref.~\cite{Bialous2016}. Recently, it was successfully applied to the measured molecular resonances in $^{166}$Er and $^{168}$Er~\cite{Mur2015}. These systems preserve \T invariance, whereas for the case of violated \T invariance in the presence of missing levels there was a lack of experimental studies. This was the motivation for the experiments presented in this Letter. 

\begin{figure}[h!]
\includegraphics[width=0.9\linewidth]{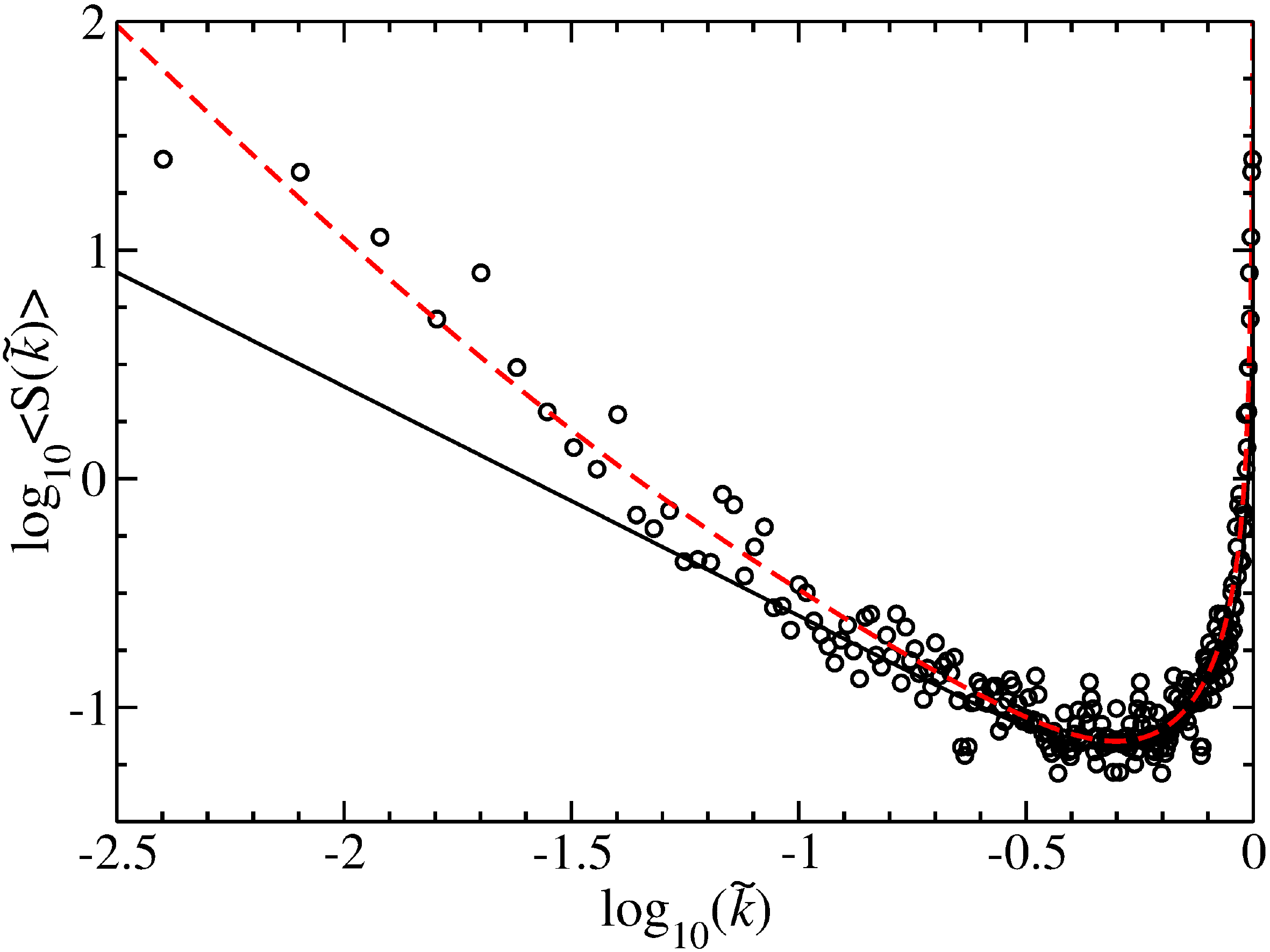}
\caption{(Color online) The average power spectrum. The experimental results (black circles) are compared to that for the eigenvalues of random matrices from the GUE (black full line) and the corresponding missing-level statistics (red [gray] dashed line). The fraction of the observed levels was unambiguously determined to $\varphi=0.965\pm 0.005$, by comparison of the experimental power spectrum to the latter. 
}\label{Fig4}
\end{figure}

In Fig.~\ref{Fig4} the experimental power spectrum (circles) is compared to that for the eigenvalues of random matrices from the GUE (black full line). Both curves are plotted versus $\tilde k$. We observe that, firstly, both curves start to deviate from each other below $\log_{10}\tilde k\lesssim -0.5$. Secondly, the experimental $\langle S(\tilde k)\rangle$ does not exhibit a clear power law behavior for small $\tilde k$. These deviations, and also those observed for the long-range spectral fluctuations in Fig.~\ref{Fig3} (c) and (d) cannot result from a mixing of symmetries~\cite{Robnik2005,Salasnich2005,Santhanam2005,Relano2008,Dietz2014}, since the short-range spectral fluctuations are well described by GUE statistics. However, similar to Ref.~\cite{Mur2015}, they can be attributed to the small fraction of missing levels, as demonstrated in the sequel.  

{\it Missing level statistics.}---
As stated above, the completeness of energy spectra is a rather rare situation in experimental investigations~\cite{Liou1972,Zimmermann1988,Agvaanluvsan2003}. The problem of missing levels can be circumvented in open systems, like microwave billiards or microwave networks, where scattering matrix elements are available. Their fluctuation properties provide measures for the chaoticity, e.g., in terms of their correlation functions~\cite{Dietz2009,Dietz2010} or the enhancement factor~\cite{Lawniczak2010,Lawniczak2015,Lawniczak2016a}. For closed systems, analytical expressions were derived for incomplete spectra based on RMT in Ref.~\cite{Bohigas2004}. The nearest-neighbor spacing distribution is expressed in terms of the $(n+1)$st nearest-neighbor spacing distribution $P(n,s)$, with $P(0,s)=P(s)$. It is well approximated by $P(n,s)\simeq\gamma s^\mu e^{-\varkappa s^2}$, where for $n=0,1$ $\mu =1,4$ for the GOE and $\mu =2,7$ for the GUE~\cite{Stoffregen1995}. The coefficients $\gamma$ and $\varkappa$ are obtained from the normalization of $P(n,s)$ to unity and the scaling of $s$ to average spacing unity, respectively. If the fraction of detected eigenvalues $\varphi$ is close to unity, the nearest-neighbor spacing distribution accounting for missing levels, is given by
\begin{equation}
p(s)\simeq P\left(\frac{s}{\varphi}\right)+(1-\varphi)P\left(1,\frac{s}{\varphi}\right)+.... 
\label{abst}
\end{equation}
Similarly, the number variance $\Sigma^2$ and the stiffness $\Delta_3$  may be expressed in terms of those for complete spectra ($\varphi =1$),
\begin{equation}
\sigma^2(L)=(1-\varphi)L+\varphi^2\Sigma^2\left(\frac{L}{\varphi}\right)
\label{sigma2}
\end{equation}
and 
\begin{equation}
\delta_3(L)=(1-\varphi)\frac{L}{15}+\varphi^2\Delta_3\left(\frac{L}{\varphi}\right).
\label{delta3}
\end{equation}
In Fig.~\ref{Fig3} the functions Eq.~(\ref{abst})-(\ref{delta3}) are plotted for $\varphi=0.965$ as red [gray] dashed lines. The agreement with the corresponding experimental results is remarkable. Like the experimental nearest-neighbor spacing distribution the curve obtained from  Eq.~(\ref{abst}) is close to that of the eigenvalues of random matrices from the GUE. This feature enabled the assignment of the GUE as the RMT model applicable to the experimental data. In order to corroborate that the deviations from GUE observed in Figs.~\ref{Fig3} and~\ref{Fig4} indeed are solely due to missing levels we analysed power spectra. An analytical expression was derived for the power spectrum of incomplete spectra in Ref~\cite{Molina2007},
\begin{eqnarray}
\label{Analytical}
\langle s(\tilde k)\rangle &=&\nonumber
\frac{\varphi}{4\pi^2}\left[\frac{K\left(\varphi\tilde k\right)-1}{\tilde k^2}+\frac{K\left(\varphi\left(1-\tilde k\right)\right)-1}{(1-\tilde k)^2}\right]\\
&+& \frac{1}{4\sin^2(\pi\tilde k)} -\frac{\varphi^2}{12},
\label{noise}
\end{eqnarray}
which for $\varphi =1$ yields that for complete spectra.
Here, $0\leq \tilde k\leq 1$ and $K(\tau)$ is the spectral form factor, which equals $K(\tau)=\tau$ for the GUE.

\begin{figure}[h!]
{\includegraphics[width=0.9\linewidth]{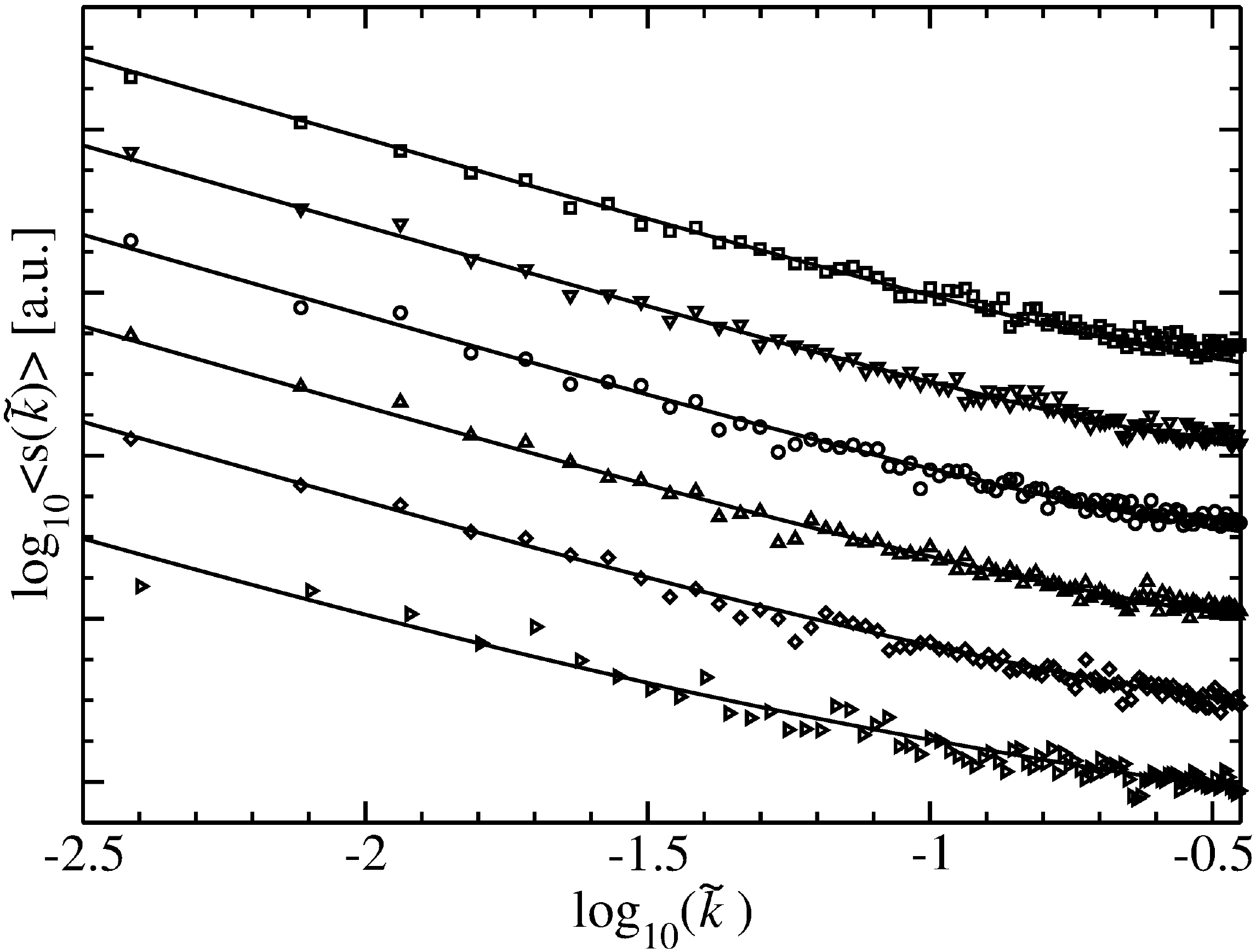}}
\caption{Illustration of the sensitivity of the average power spectrum to changes in the fraction of observed levels $\varphi$. The symbols show results which were generated from the experimental data by randomly eliminating resonance frequencies. Triangles-right show the data from~\reffig{Fig4} with $\varphi=0.965$, diamonds correspond to $\varphi=0.9$, triangles-up to $\varphi=0.85$, circles to $\varphi=0.8$, triangles-down to $\varphi =0.75$ and squares to $\varphi=0.7$. In order to allow a distinct demonstration of the excellent agreement between the theoretical and the experimental results, all except the curve for $\varphi=0.965$ were shifted with respect to their lower neighbor by unity. 
}\label{Fig5}
\end{figure}

This analytical result is shown as red dashed curve in Fig.~\ref{Fig4}. The fraction of observed levels, actually, was determined to $\varphi=0.965\pm 0.005$ from the power spectrum, which depends particularly sensitively on the value of $\varphi$. This is illustrated in Fig.~\ref{Fig5}, where we compare its asymptotic behavior to experimental results. Here, the fraction $\varphi$ was varied by randomly eliminating resonance frequencies. The power spectra for different values of $\varphi$ lie close to one another. However, they still are clearly distinguishable. To illustrate this, each curve was shifted by unity with respect to its lower neighbor in Fig.~\ref{Fig5}. Even for the case of only 70~$\%$ of observed levels, we find good agreement between the analytical result Eq.~(\ref{noise}) and the experimental one. 

{\it Conclusions.}---
We present first experimental studies of the fluctuation properties in incomplete spectra of microwave networks simulating chaotic quantum graphs with broken time reversal symmetry. The experimental results are in good agreement with the analytical expressions for missing level statistics Eqs.~(\ref{abst})-(\ref{delta3}) derived in Ref.~\cite{Bohigas2004} and Eq.~(\ref{noise}) for the power spectrum given in Ref.~\cite{Molina2007}. All these expressions explicitly take into account the fraction of observed levels $\varphi$, however, the power spectrum is particularly sensitive to it. Therefore, we used it to determine the fraction of observed levels, $\varphi=0.965\pm 0.005$, in the experimental spectra. The symmetry (GUE) of the system was determined from the nearest-neighbor spacing distribution which depends only weakly on $\varphi$ for $\varphi\gtrsim 0.9$. Long-range spectral fluctuations were then used to confirm this assignment. The excellent agreement between the experimental and the analytical results, demonstrated in Fig.~\ref{Fig5} for a range of $0.7\leq\varphi\leq 0.965$, clearly proves the vigorousness of the power spectrum for the description of incomplete spectra of quantum systems with violated \T invariance. 

This work was partially supported by the Ministry of Science and Higher Education grant UMO-2013/09/D/ST2/03727 and the EAgLE project (FP7-REGPOT-2013-1, Project Number: 316014).


\begin{thebibliography}{99}

\bibitem{Gomez2011} J. M. G. G\'omez, K. Kar, V. K. B. Kota, R. A. Molina, A. Rela\~no, and J. Retamosa, Phys. Rep. {\bf 499}, 103 (2011).

\bibitem{Weidenmueller2009} H. A. Weidenm\"uller and G. E. Mitchell, Rev. Mod. Phys. {\bf 81}, 539 (2009).

\bibitem{Haake2001} F. Haake, {\it Quantum Signatures of Chaos} (Springer-Verlag, Heidelberg, 2001).

\bibitem{Berry1977} M.V. Berry and M. Tabor, Proc. R. Soc. A {\bf 356}, 375 (1977).

\bibitem{Footnote} The Gaussian ensembles are ensembles of random matrices, of which the entries are Gaussian distributed with zero mean. Random matrices from the GOE and the GUE are real symmetric and hermitian, respectively. 

\bibitem{Mehta1990} M.~L. Mehta, {\it Random Matrices} (Academic Press, London, 1990).

\bibitem{Bohigas1984} O. Bohigas, M.~J. Giannoni, and C. Schmit, Phys. Rev. Lett. {\bf 52}, 1 (1984).

\bibitem{Stoeckmann2000} H.-J. St{\"o}ckmann, \emph{Quantum Chaos: An Introduction} (Cambridge University Press,Cambridge,2000)

\bibitem{Dietz2015} B. Dietz and A. Richter, CHAOS {\bf 25}, 097601 (2015).

\bibitem{Hul2004} O. Hul, S. Bauch, P. Pako\'nski, N. Savytskyy, K. \.Zyczkowski, and L. Sirko, Phys. Rev. E {\bf 69}, 056205 (2004).

\bibitem{Lawniczak2010}M. {\L}awniczak,  S. Bauch, O. Hul, and L. Sirko, Phys. Rev. E {\bf 81}, 046204 (2010).

\bibitem{Kottos1997} T. Kottos, U. Smilansky, Phys. Rev. Lett. {\bf 79}, 4794 (1997).

\bibitem{Kottos1999} T. Kottos, U. Smilansky, Ann. Phys. {\bf 274}, 76 (1999).

\bibitem{Pakonski2001} P. Pako\'nski, K. \.Zyczkowski, M. Ku\'s\, J. Phys. A {\bf 34} 9303 (2001).

\bibitem{Pauling1936} L. J. Pauling, Chem. Phys. {\bf 4}, 673 (1936).

\bibitem{Sanchez1988} J. A. Sanchez-Gil, V. Freilikher, I. Yurkevich, and A.~A. Maradudin, Phys. Rev. Lett. {\bf 80}, 948 (1998).

\bibitem{Mittra1971} R. Mittra, S. W. Lee, {\it Analytical Techniques in the Theory of Guided Waves} (Macmillan, NY, 1971).

\bibitem{Kowal1990} D. Kowal, U. Sivan, O. Entin-Wohlman, and Y. Imry, Phys. Rev. B {\bf 42}, 9009 (1990).

\bibitem{Imry1996} Y. Imry, {\it Introduction to Mesoscopic Systems} (Oxford, NY, 1996).

\bibitem{Gnutzmann2004} S. Gnutzmann and A. Altland, Phys. Rev. Lett. {\bf 93}, 194101 (2004).

\bibitem{Pluhar2014} Z. Pluha\v r and H. A. Weidenm\"uller, Phys. Rev. Lett. {\bf 112}, 144102 (2014).

\bibitem{Lawniczak2008} M. {\L}awniczak, O. Hul, S. Bauch, P. Seba, and L. Sirko, Phys. Rev. E {\bf 77}, 056210 (2008).

\bibitem{Lawniczak2011} M. {\L}awniczak,  S. Bauch, O. Hul, and L. Sirko, Phys. Scr. {\bf T143},  014014 (2011).

\bibitem{Hul2012} O. Hul, M.~{\L}awniczak, S. Bauch, A. Sawicki, M. Ku\'s, L. Sirko, Phys. Rev. Lett {\bf 109}, 040402 (2012).

\bibitem{Lawniczak2014} M.~{\L}awniczak, A.~Sawicki, S.~Bauch, M.~Ku\'s, and L.~Sirko, Phys. Rev E {\bf 89,} 032911 (2014).

\bibitem{Allgaier2014} M. Allgaier, S. Gehler, S. Barkhofen, H.-J. St\"ockmann, and U. Kuhl, Phys. Rev. E {\bf 89}, 022925 (2014).

\bibitem{Bohigas1983} O. Bohigas, R. U. Haq, and A. Pandey, in {\it Nuclear Data for Science and Technology}, ed. by K. H. B\"ockhoff (Reidel, Dordrecht, 1983).

\bibitem{Sieber1993} M. Sieber, U. Smilansky, S. C. Creagh, and R. G. Littlejohn, J. Phys. A {\bf 26}, 6217 (1993).

\bibitem{Dietz2014} B. Dietz, T. Guhr, B. Gutkin, M. Miski-Oglu, and A. Richter, Phys. Rev. E {\bf 90}, 022903 (2014).

\bibitem{French1985} J. B. French, V. K. B. Kota, A. Pandey, and S. Tomsovic, Phys. Rev. Lett. {\bf 54}, 2313 (1985).

\bibitem{Mitchell2010} C. E. Mitchell, A. Richter, H. A. Weidenm\"uller, Rev. Mod. Phys. {\bf 82}, 2845 (2010).

\bibitem{Pluhar1995} Z. Pluha\v r, H. A. Weidenm\"uller, J. A. Zuk, C. H. Lewenkopf, and F. J. Wegner, Ann. Phys. {\bf 243}, 1 (1995).

\bibitem{Dietz2009} B. Dietz, T. Friedrich, H. L. Harney, M. Miski-Oglu, A. Richter, F. Sch\"afer, J. Verbaarschot, and H. A. Weidenm\"uller, Phys. Rev. Lett. {\bf 103}, 064101 (2009).

\bibitem{Dietz2010} B. Dietz, T. Friedrich, H. L. Harney, M. Miski-Oglu, A. Richter, F. Sch\"afer and H. A. Weidenm\"uller, Phys. Rev. E {\bf 81}, 036205 (2010).

\bibitem{So1995} P. So, S. M. Anlage, E. Ott, and R. N. Oerter, Phys. Rev. Lett. {\bf 74}, 2662 (1995).

\bibitem{Stoffregen1995} U. Stoffregen, J. Stein, H.-J. St\"ckmann, M. Ku\'s, and F. Haake, Phys. Rev. Lett. {\bf 74}, 2666 (1995).

\bibitem{Wu1998} J. S. A. Bridgewater, A. Gokirmak, and S. M. Anlage, Phys. Rev. Lett. {\bf 81}, 2890 (1998).
\bibitem{Frisch2014} A. Frisch, M. Mark, K. Aikawa, F. Ferlaino, J. Bohn, C. Makrides, A. Petrov, and S. Kotochigova, Nature (London) {\bf 507}, 475 (2014).

\bibitem{Mur2015} J. Mur-Petit and R. A. Molina, Phys. Rev. E {\bf 92}, 042906 (2015). 

\bibitem{Liou1972} H.~I. Liou, H.~S. Camarda, and F. Rahn, Phys. Rev. C {\bf 5}, 131 (1972). 

\bibitem{Zimmermann1988} T. Zimmermann, H. K\"oppel, L.~S. Cederbaum, G. Persch, and W. Demtr\"oder, Phys. Rev. Lett. {\bf 61}, 3 (1988).

\bibitem{Agvaanluvsan2003} U. Agvaanluvsan, G.~E. Mitchell, J.~F. Shriner Jr., M. Pato, Phys. Rev. C {\bf 67}, 064608 (2003). 

\bibitem{Bohigas2004} O. Bohigas and M. P. Pato, Phys. Lett. B {\bf 595}, 171 (2004).

\bibitem{Molina2007} R.A. Molina, J. Retamosa, L. Mu\~noz, A. Rela\~no, and E. Faleiro, Phys. Lett. B {\bf 644}, 25 (2007).

\bibitem{Relano2002} A. Rela\~no, J.M.G. G\'omez, R. A. Molina, J. Retamosa, and E. Faleiro, Phys. Rev. Lett. {\bf 89}, 244102 (2002).

\bibitem{Faleiro2004} E. Faleiro, J. M. G. G\'omez, R. A. Molina, L. Mu\~noz, A. Rela\~no, and J. Retamosa, Phys. Rev. Lett. {\bf 93}, 244101 (2004).

\bibitem{Jones} D. S. Jones, {\it Theory of Electromagnetism} (Pergamon Press, Oxford, 1964), p. 254.

\bibitem{Savytskyy2001} N. Savytskyy, A. Kohler, S. Bauch, R. Bl\"umel, and L. Sirko, Phys. Rev. E {\bf 64}, 036211 (2001).

\bibitem{Robnik2005} J. M. G. G\'omez, A. Rela\~no, J. Retamosa, E. Faleiro, L. Salasnich, M. Vrani\v car, and M. Robnik, Phys. Rev. Lett. {\bf 94}, 084101 (2005).

\bibitem{Salasnich2005} L. Salasnich, Phys. Rev. E {\bf 71}, 047202 (2005).

\bibitem{Santhanam2005} M. S. Santhanam and J. N. Bandyopadhyay,  Phys. Rev. Lett. {\bf 95}, 114101 (2005).

\bibitem{Relano2008} A. Rela\~no, Phys. Rev. Lett. {\bf 100}, 224101 (2008).

\bibitem{Faleiro2006} E. Faleiro, U. Kuhl, R.A. Molina, L. Mu\~noz, A. Rela\~no, and  J. Retamosa, Phys. Lett. A {\bf 358}, 251 (2006).

\bibitem{Bialous2016}M. Bia{\l}ous, V. Yunko, M. {\L}awniczak, S. Bauch, B. Dietz, and L. Sirko, in preparation.

\bibitem{Lawniczak2016a} M. {\L}awniczak, S. Bauch, and L. Sirko, in {\it Handbook of Applications of Chaos Theory}, eds. Christos Skiadas and Charilaos Skiadas (CRC Press, Boca Raton, USA, 2016), p. 559.

\bibitem{Lawniczak2015} M. {\L}awniczak, M. Bia{\l}ous, V. Yunko, S. Bauch, and L. Sirko, Phys. Rev. E {\bf 91}, 032925 (2015).

\end{thebibliography}
\end{document}